# Study of antiferromagnetic and nematic phase transitions in BaFe$_2$(As$_{1-x}$P$_x$)$_2$ by AC micro-calorimetry and SQUID magnetometry

X. Luo[1,2,†], V. Stanev[1,3], B. Shen[1], L. Fang[1], X. S. Ling[2], R. Osborn[1], S. Rosenkranz[1], W.-K. Kwok[1], U. Welp[1,†]

[1] Materials Science Division, Argonne National Laboratory, Argonne, IL 60439, USA

[2] Department of Physics, Brown University, Providence, RI 02912, USA

[3] Department of Physics, University of Maryland, College Park, MD 20742, USA

We study the antiferromagnetic (AFM) and structural phase transitions in single crystal BaFe$_2$(As$_{1-x}$P$_x$)$_2$ ( $x = 0$, $0.3$ ) at temperatures $T_N$ and $T_S$, respectively, by high resolution ac microcalorimetry and SQUID magnetometry. The specific heat measurements of both as grown and annealed BaFe$_2$As$_2$ displays a sharp peak at the AFM/Structural transitions. A kink in the entropy of annealed BaFe$_2$As$_2$ gives evidence for splitting of the two transitions by approximately 0.5K. No additional features could be identified in the specific heat of both BaFe$_2$As$_2$ and BaFe$_2$(As$_{0.7}$P$_{0.3}$)$_2$ in the temperature regions around $T^* > T_S$ where torque measurements [S. Kasahara et al., Nature **486**, 382 (2012)] had revealed the "true" nematic phase transition, indicating that the behavior at $T^*$ does not represent a 2$^{nd}$ order phase transition, and that the phase transition of BaFe$_2$(As$_{1-x}$P$_x$)$_2$ into the orthorhombic phase does occur at $T_S$.

One of the key issues in understanding Iron-based superconductors (FeSCs)[1] lies in the peculiar normal state properties of these materials. Most of the parent compounds of FeSCs develop an antiferromagnetic (AFM) stripe order spin density wave (SDW) ground state below a phase transition temperature $T_N$. This AFM transition is suppressed by doping which eventually leads to superconductivity. What is particularly interesting and sets FeSCs apart from other unconventional superconductors is that the suppression of the AFM parent state by doping is preceded or coincident with a structural transition of the lattice from tetragonal to orthorhombic symmetry[2]. The interplay of the magnetic and structural transitions generates rich physics. In the case of 1111 materials (RFeAsO, R = rare earth), the two transitions are split ($T_S > T_N$) and of second order.[3] However, in 122 materials ($AFe_2As_2$, A = alkaline-earth), the situation is more complicated. While neutron scattering suggests simultaneous weakly first order AFM and structural transitions over the entire phase diagram of hole doped $Ba_{1-x}K_xFe_2As_2$ [4], transport, specific heat, magnetization, and diffraction experiments indicate split AFM and structural transitions in electron doped $Ba(Fe_{1-x}Co_x)_2As_2$ [5,6,7], with evidence of the existence of a tricritical point separating first and second order magnetic transitions[6,7].

Although a conventional phonon (lattice vibration) driven mechanism of the structural transition cannot be ruled out completely, this transition has generally been considered as a manifestation of electronic nematic order [8], which has also been inferred from the unusual anisotropy in resistivity[9,10], optical conductivity[11] and orbital occupancy[12] observed at temperatures above the structural transition. The origin of nematic order has been ascribed to either a spontaneous ferro-orbital order with unequal occupations between the Fe $d_{xz}$ and $d_{yz}$ orbitals[13-17] or an Ising spin-nematic order where the $Z_2$ symmetry between the two possible SDW ordering wave vectors $\mathbf{Q_1} = (0, \pi)$ and $\mathbf{Q_2} = (\pi, 0)$ in the 1-Fe Brillouin Zone (BZ) is broken before the O(3) spin rotational symmetry[18-22]. Regardless of the exact microscopic origin of nematicity, a phenomenological treatment of the problem based on Ginzburg-Landau (GL) theory yields a good description of the order of the AFM and structural transitions and the possibility of a tricritical point in the phase diagram[7, 8, 23].

Recent magnetic torque measurements on BaFe$_2$(As$_{1-x}$P$_x$)$_2$ [24] and EuFe$_2$(As$_{1-x}$P$_x$)$_2$ [25] single crystals under in-plane magnetic field rotation reveal the onset of two fold oscillations which break the tetragonal symmetry at a temperature $T^*$ well above (>30K) the commonly accepted nematic/structural transition at $T_S$. These results were interpreted [24,25] as signature of a "true" 2$^{nd}$ order nematic phase transition at $T^*$ leading from the high-temperature tetragonal phase to a low-temperature phase with C$_2$-symmetry whereas the conventional structural transition at $T_s$ ceases to be a true phase transition but is regarded as a meta-nematic transition. This "true" transition at $T^*$ is found to persist even for doping levels in the nonmagnetic superconducting regime, which dramatically changes the phase diagrams of BaFe$_2$(As$_{1-x}$P$_x$)$_2$ and EuFe$_2$(As$_{1-x}$P$_x$)$_2$. For instance, consideration needs to be given to the number of degrees of freedom required for stabilizing a nematic state over such a wide temperature range [26] in a macroscopically tetragonal lattice. Measurements of the strain dependent resistivity anisotropy[10] or of the shear elastic constants [27] of BaFe$_2$As$_2$ (parent compound) do not yield evidence for additional phase transitions above the usual structural transition. Similarly, a transition at $T^*$ in EuFe$_2$(As$_{1-x}$P$_x$)$_2$ has not been noted in previous thermopower measurements [28]. Magnetic torque however is directly related to the spin nematic order parameter [21] possibly facilitating the observation of a nematic phase transition. A recent STM/STS study on NaFeAs single crystals [29] also revealed the persistence of local electronic nematicity up to temperatures of almost twice $T_S$. In this case, residual strains in the sample in conjunction with a large nematic susceptibility were considered as possible origin of such symmetry breaking. Thus, the question whether phenomena at $T^*$ represent a $2^{nd}$ order phase transition, a cross-over associated with the onset of sizable short-range correlations and fluctuations, or spurious effects due to frozen-in strains remains unresolved.

Here we present a study of single crystal $BaFe_2(As_{1-x}P_x)_2$ by high resolution ac micro-calorimetry [30] and SQUID magnetometry in an attempt to investigate the various phase transitions and to explore the "true" nematic phase transition. A $2^{nd}$ order nematic transition should appear in the thermal channel, i.e., in the specific heat. Specific heat is a direct probe of thermodynamic phase transitions; it does not require the application of external fields such as strain or magnetic field, which could break the symmetry. As the sample is significantly thicker than the supporting $Si_3N_4$-membrane of the calorimeter (see below), the effects resulting from strains due to differential thermal contraction are negligible. Furthermore, the specific heat is independent of the degree of twinning in the sample. Results from our specific heat measurements of both as grown and annealed $BaFe_2As_2$ reveal a sharp peak at the AFM/Structural transitions. A kink in the entropy of annealed $BaFe_2As_2$ gives evidence for splitting of the two transitions, with the $2^{nd}$ order structural transition preceding the AFM transition by approximately 0.5 K. No additional features could be identified in the specific heat of both $BaFe_2As_2$ and $BaFe_2(As_{0.7}P_{0.3})_2$ in the temperature regions where torque measurements [24] revealed the nematic phase transition eventhough the Ginzburg-Landau model used to fit the magnetic torque data indicates that the expected thermal anomaly is easily within our experimental resolution. Similarly, magnetization measurements of as grown and annealed $BaFe_2As_2$ show sharp steps at the AFM/structual transition while no evidence for another phase transition could be found.

High quality BaFe$_2$(As$_{1-x}$P$_x$)$_2$ crystals were grown by the self-flux method as described elsewhere [31]. Annealing of as grown BaFe$_2$As$_2$ was carried out in an evacuated quartz tube together with BaAs flux at 800 $^o$C for 72 hours. [32] High resolution specific heat measurements were performed with a home built membrane-based ac microcalorimeter. The calorimeter utilizes a pair of micro-fabricated Au-1.7%Co and Cu thermocouples as the temperature sensor on top of a 150-nm-thick Si$_3$N$_4$ membrane. Accurate calibration of the calorimeter was accomplished by zero-field and in-field measurements of a Au standard sample, which has a heat capacity comparable to our samples. Single crystal samples of BaFe$_2$As$_2$, with dimensions of ~ $120 \times 110 \times 20$ $\mu m^3$ for the as grown and ~ $130 \times 180 \times 13$ $\mu m^3$ for the annealed sample, respectively, are mounted onto the calorimeter with a minute amount of Apiezon N grease. An ac heater current at a typical frequency of 20.5 Hz is adjusted so as to induce ~200 mK oscillations in the sample temperature. The magnetization measurement were performed in a commercial SQUID magnetometer. The samples were mounted to a quartz fiber sample holder in order to minimize background signals and artifacts arising from the thermal expansion of the sample holder.

The main panel of Fig. 1 shows the temperature dependence of the specific heat for both as grown and annealed BaFe$_2$As$_2$ samples. As has been observed previously [33], the sharpness of the transition and the transition temperature clearly increase upon annealing. For the annealed sample, we observe a transition temperature of 137 K and a specific heat peak width (FWHM) of 0.7 K. The corresponding values of the as-grown sample are 133 K and 1.2 K, respectively, whereas the BaFe$_2$As$_2$ sample used in torque measurements [24] had a transition temperature of ~ 135 K.

Integrating $C/T$ over temperature yields the change in entropy of $BaFe_2As_2$ across the transition as shown in the inset of Fig. 2. A clear step-like anomaly is discernable at the AFM/Structural transitions of both samples. The detailed shape of the anomaly is shown in the main panel obtained by subtracting a normal state background from the original entropy, indicated by the dashed lines in the inset. The change in entropy at the transition, extracted by approximating the transition as a sharp step, is found to be ~0.5 J/mol K for both as grown and annealed $BaFe_2As_2$. This value is slightly smaller than that from a previous report [33], where a change of entropy of ~0.84 J/mol K was found for an annealed crystal with a transition temperature of 140 K. The change in entropy across the AFM transition is substantially smaller than the value of $R\ln(2)$ expected for the onset of long-range magnetic order in a $S=1/2$ – system, indicative of pronounced magnetic fluctuations [34]. The shape of the $C/T$- and S-curves, particularly of the as-grown sample, is consistent with a broadened 1st order transition as well as with a 2nd order magnetic transition accompanied by critical fluctuations [35]. However, for our annealed sample a clear kink in $S(T)$ is seen near the top of the transition (shown by the black arrow in Fig. 2) about 0.5 K above the peak temperature in the specific heat (position of the double headed arrow in Fig. 2), followed by a tail towards high temperatures. Such behavior does not arise due to critical fluctuations, but might be evidence that the structural transition and AFM transition in annealed $BaFe_2As_2$ are actually split, with the 2nd order structural transition about 0.5 K above the 1st order AFM transition. Similar results have also been obtained recently from high resolution X-ray diffraction measurements on as-grown $BaFe_2As_2$ [7], showing that the 2nd order structural transition and the 1st order AFM transition are in fact separated by approximately 0.75 K.

We also measured the specific heat of a near-optimum doped $BaFe_2(As_{1-x}P_x)_2$ (x=0.3) crystal with dimensions of $113\times154\times22$ $\mu m^3$. Figure 3 shows the heat capacity measurements from the same calorimeter used for the parent compound. A small step like feature is found near 29K, indicative of the superconducting transition of the sample shown in more detail in the upper inset after subtraction of a normal state background.

The inset of Fig. 1 and the lower inset of Fig. 3 show the specific heat on largely expanded scales after subtraction of smooth polynomial backgrounds. Within our resolution of $10^{-4}$, no feature can be identified that would indicate a phase transition near the expected nematic transition temperatures of 170 K and 90 K of the parent compound and optimally doped sample, respectively.

We evaluate the expected specific heat signature at the nematic transition using the Ginzburg-Landau free energy for BaFe$_2$As$_2$ as given in Ref. 24:

$$F[\delta,\psi] = \left[t_s \delta^2 - u\delta^4 + v\delta^6\right] + \left[t_p \psi^2 + w\psi^4\right] - g\delta\psi$$

Here $\delta = \dfrac{a-b}{a+b}$ denotes the lattice distortion and $\psi$ is the nematic order parameter. $t_{s,p} = \dfrac{T - T_{s,p}^{(0)}}{T_{s,p}^{(0)}}$ is the reduced temperature of the structural/nematic transitions, with $T_{s,p}^{(0)}$ denoting the transition temperatures in the absence of coupling between the two order parameters, i.e. $g = 0$. The coefficients $u$, $v$, and $w$ are determined in Ref. 24 from fits to the torque and XRD data on a BaFe$_2$As$_2$ crystal with a transition temperature very close to the one investigated here. This Ginzburg-Landau model yields a 2$^{nd}$ order nematic phase transition at $T^* > T_p^{(0)}$ and a meta-nematic transition at $T_s > T_s^{(0)}$. By using the same model, we derive the temperature dependence of the free energy $F(T)$, entropy $S(T)$ and specific heat $C(T)$. The latter two are shown in Fig. 4. The theoretical curves reproduce the shape of the experimental curves quite well regarding the AFM/structural transition, with a similar sharp peak in the specific heat and a step in the entropy at $T_s$, though the experimental entropy curve is more smeared out possibly due to fluctuations or inhomogeneity in the sample. In addition, the theoretical specific heat curve also reveals a small step at the nematic transition ($T^*$). In order to evaluate the expected height of this step, we consider the ratio of the change in entropy at $T_s$, as given by $\Delta S = -\left( \partial F/\partial T|_{T_s^+} - \partial F/\partial T|_{T_s^-} \right)$, and the step in the specific heat at $T^*$, $\Delta C = -T \partial^2 F/\partial T^2|_{T^*}$. This ratio is independent of an over-all scale factor and is found from the GL model to be $\Delta S|_{T_s}/\Delta C|_{T^*} \approx 5$. From Fig. 2 we obtain the change in entropy at the AFM/structural transition of 0.5 J/mol K, yielding the expected height of the specific heat anomaly at $T^*$ of ~ 0.1 J/mol K. Considering that the noise level at ~170 K (the expected $T^*$ for BaFe$_2$As$_2$) is ~0.012 J/mol K, we should be able to distinguish such a feature, indicating that there are no 2$^{nd}$ order phase transition at $T^*$ and that the transition into the C$_2$-phase occurs at $T_S$.

The magnetization of both as grown and annealed BaFe$_2$As$_2$ samples were measured in an applied field of 1 T along the basal plane and along the c-axis, respectively. The results are summarized in Fig. 5. A sharp step-like feature in the magnetization for both applied field directions in as grown and annealed samples indicates the AFM/Structural transition. The transition temperatures are consistent with those obtained from the specific heat measurements. The value of the magnetization and the drop at $T_N$ for $H \parallel ab$ are higher than that for $H \parallel c$ by a factor of ~2-3, consistent with the in-plane spin arrangements in the Fe-As planes [2]. Above the magneto-structural transition the magnetization increases linearly with temperature [36], distinctly different from the temperature-independent Pauli paramagnetism of itinerant carriers as well as the 1/T-decrease in the Curie-Weiss law of independent local moments. Such linear temperature dependence has been reported previously for several iron-based superconductors, including BaFe$_2$As$_2$ [37], CaFe$_2$As$_2$ [38], LaFeAsO$_{1-x}$F$_x$ [39], Ca(Fe$_{1-x}$Co$_x$)$_2$As$_2$ [39] and SrFe$_2$As$_2$ [40], as well as high-$T_c$ La$_{2-x}$SrCuO$_{4-y}$ [41]. It was suggested to be a consequence of strong AFM correlations [42, 43] persisting in the paramagnetic state or, alternatively, of flat electronic bands caused by the quasi 2D crystal structure [44]. Subtraction of the aforementioned linear $M(T)$ background from the raw data yields a detailed presentation of the magnetic transition shown in the main panel of Fig. 3. As can be seen, the transition is slightly sharper for the annealed compound. Specifically, the broadening right above the transition found in the as grown sample almost disappears after annealing. Such a sharp transition without any indication of precursors is quite unexpected if magnetic fluctuations play a key role in the magnetostructural transition. However, this seeming contradiction can be explained by the fact that the uniform magnetization is mostly sensitive to fluctuations at $Q = 0$ in the BZ, and is therefore, not a direct measurement of the fluctuations at the SDW ordering wave vectors ($Q = (0, \pi)$ and $(\pi, 0)$ ). Recently, a scaling relation between the NMR spin lattice relaxation and the elastic shear modulus in Ba(Fe$_{1-x}$Co$_x$)$_2$As$_2$, was discovered [45], indicative of strong coupling between magnetic and structural fluctuations.

In summary, we presented SQUID magnetometry and high resolution ac microcalorimetry measurements of single crystal BaFe$_2$(As$_{1-x}$P$_x$)$_2$ ( $x = 0$, 0.3 ). Results on both as grown and annealed BaFe$_2$As$_2$ reveal a sharp peak at the AFM/Structural transitions. A kink in the entropy of annealed BaFe$_2$As$_2$ gives evidence for splitting of the two transitions, with the 2$^{nd}$ order structural transition preceding the AFM transition by approximately 0.5 K. Our measurements show no additional features in the specific heat of both BaFe$_2$As$_2$ and BaFe$_2$(As$_{0.7}$P$_{0.3}$)$_2$ in the temperature regions of the purported "true" nematic phase transition reported in torque measurements [24], eventhough the Ginzburg-Landau model used to fit the magnetic torque data indicates that the expected thermal anomaly should be easily observable with our experimental resolution of $10^{-4}$. We thus conclude that the behavior previously reported [24] for BaFe$_2$As$_2$ at $T^*$ does not represent a 2$^{nd}$ order phase transition, and that the phase transition into the orthorhombic phase does occur at $T_S$.

**Acknowledgement**:

We would like to thank R. M. Fernandes, F. Hardy and A. E. Böhmer for fruitful discussions. We thank T. Benseman (Materials Science Division) and R. Divan (Center for Nanoscale Materials) for help with the calorimeter fabrication. Use of the Center for Nanoscale Materials was supported by the U. S. Department of Energy, Office of Science, Office of Basic Energy Sciences, under Contract No. DE-AC02-06CH11357. This work was supported by the U.S. Department of Energy, Office of Science, Basic Energy Sciences, Materials Sciences and Engineering Division.

†To whom correspondence should be addressed.

E-mail: xluo@anl.gov and welp@anl.gov

**Figure captions:**

FIG. 1. Temperature dependence of the specific heat of as-grown and annealed BaFe$_2$As$_2$ single crystals. Inset shows the specific heat of annealed BaFe$_2$As$_2$ after a smooth polynomial background subtraction for the temperature region above the peak. Red and green curves correspond to warming and cooling runs, respectively. The dashed lines indicate the level of the anomaly expected on the basis of the GL-model. Data are off-set by 0.2 J/mol K for clarity of presentation.

FIG. 2. Temperature dependence of the entropy of as grown and annealed BaFe$_2$As$_2$, after subtraction of a smooth normal state background indicated by the dashed lines in the inset, respectively. The data for the annealed sample is shifted downward slightly to assist the eye. The dashed lines and double headed arrows demonstrate the construction used for extracting the entropy steps at the transitions. The small black arrow indicates the position of the kink in the entropy of the annealed BaFe$_2$As$_2$. Inset shows the entropies before background subtractions, with the blue and red arrows indicating the transition temperatures.

FIG. 3. Temperature dependence of the heat capacity of a BaFe$_2$(As$_{0.7}$P$_{0.3}$)$_2$ single crystal. Upper inset is a magnification of the SC transition region after subtraction of a normal state background from the original data. Lower inset is a magnification of the temperature region where the nematic transition is expected to occur. The level of resolution is about $10^{-4}$.

FIG. 4. Temperature dependence of the specific heat of BaFe$_2$As$_2$ as derived from the GL theory based numerical calculations. Insets show the temperature dependence of entropy near the AFM/structural transition.

FIG. 5. Temperature dependence of the magnetization (inset) and magnetization after subtraction of a linear background (main panel) of as grown and annealed BaFe$_2$As$_2$ in an applied field of 1T along the ab plane and c-axis.

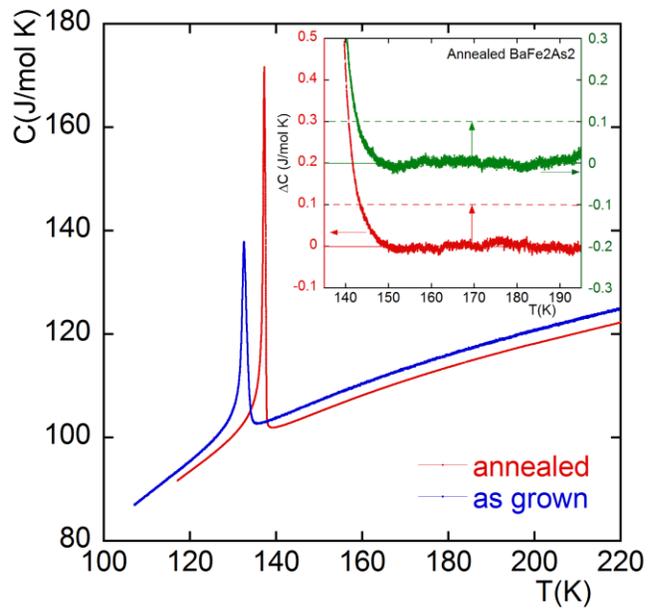

Fig. 1

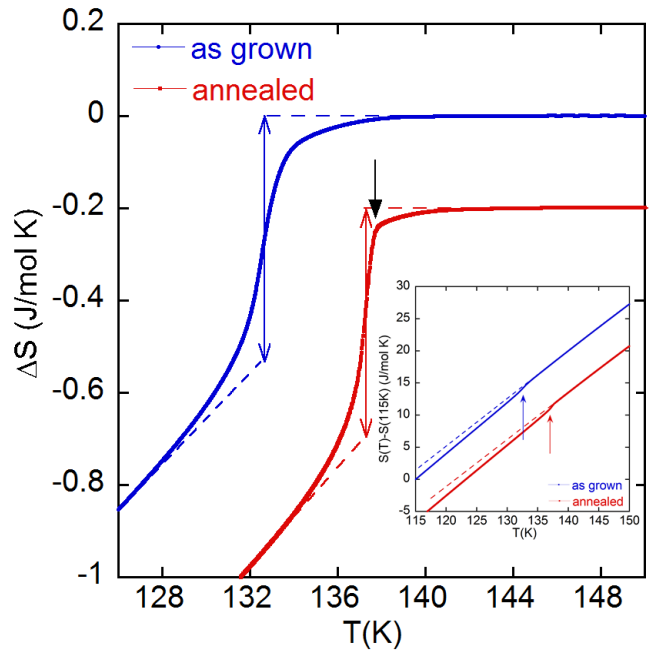

Fig. 2

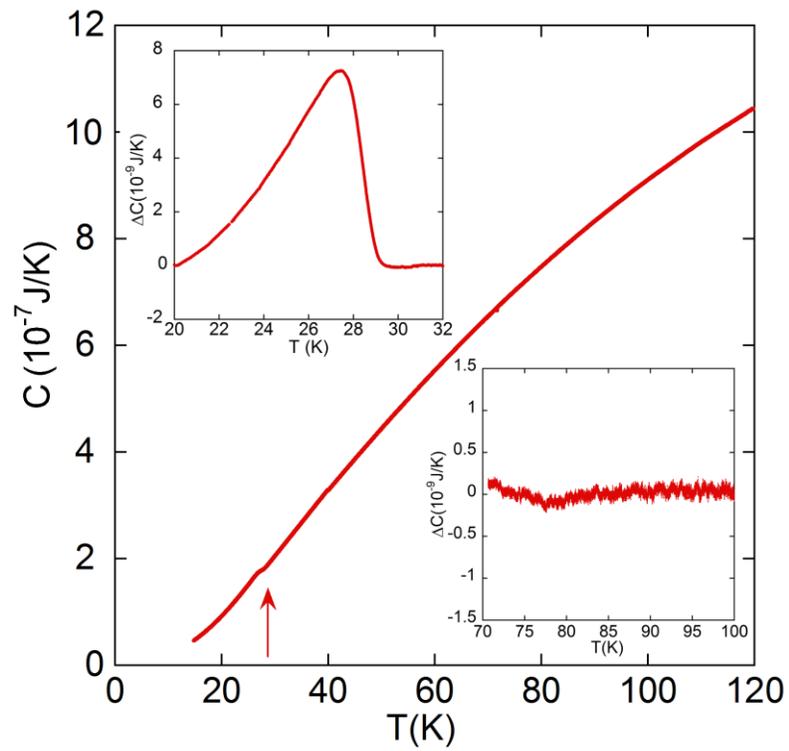

Fig. 3

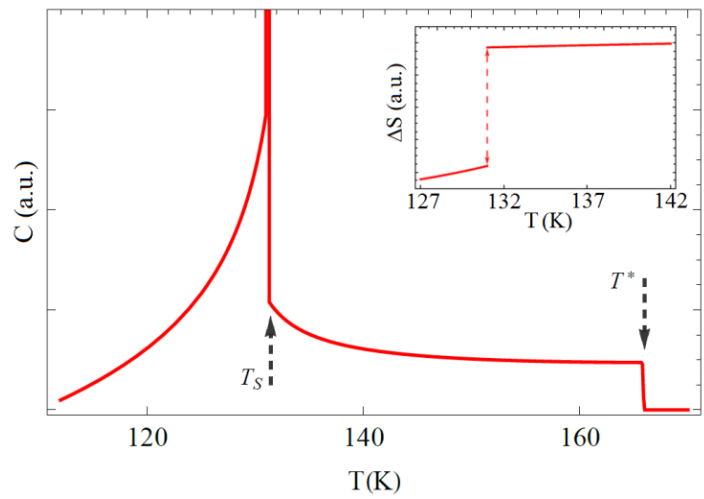

Fig. 4

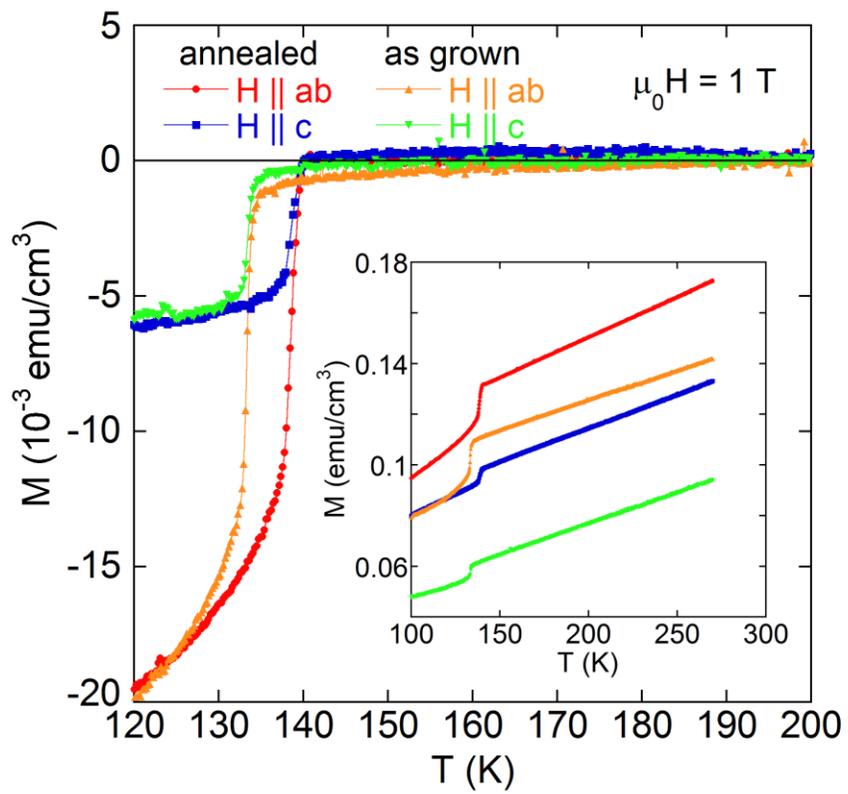

Fig. 5